\documentclass[12pt,preprint]{emulateapj}

\shorttitle{High-Metallicity Host of GRB 020819}
\shortauthors{Levesque et al.}
\slugcomment{DRAFT \today}

\begin{document}

\title{A High-Metallicity Host Environment for the Long-Duration GRB 020819}
\author{Emily M. Levesque$^{1,2}$, Lisa J. Kewley$^1$, John F. Graham$^3$, Andrew S. Fruchter$^3$}
\email{emsque@ifa.hawaii.edu, kewley@ifa.hawaii.edu, graham@stsci.edu, fruchter@stsci.edu}

\footnotetext[1]{Institute for Astronomy, University of Hawaii, 2680 Woodlawn Dr., Honolulu, HI 96822}
\footnotetext[2]{Predoctoral Fellow, Smithsonian Astrophysical Observatory}
\footnotetext[3]{Space Telescope Science Institute, 3700 San Martin Dr., Baltimore, MD 21218}

\begin{abstract}
We present spectroscopic observations of the host galaxy and explosion site of long-duration gamma-ray burst GRB 020819. We determine a metallicity for this host environment of log(O/H) + 12 = 9.0 $\pm$ 0.1, by far the highest metallicity determined for a long-duration GRB to date. We compare the metallicity and other properties of the GRB 020819 host environment to existing observations of long-duration GRB host galaxies, and consider the implications that this metallicity has for our understanding of long-duration GRB progenitor scenarios. We also consider how this unusually high metallicity may relate to the status of GRB 020819 as a ``dark" burst, with no detected optical afterglow.
\end{abstract}

\section{Introduction}
Long-duration gamma-ray bursts (LGRBs), with a duration of $>$2 s, are typically associated with the core-collapse of rapidly-rotating massive stars (Woosley 1993). These events are commonly accompanied by multi-wavelength afterglows and broad-lined Type Ic supernovae (e.g. Galama et al.\ 1998, Stanek et al.\ 2003, Woosley \& Bloom 2006). Several recent studies have presented evidence that LGRBs may occur preferentially in low-metallicity environments, a conclusion supported by both inferred (e.g. Fruchter et al.\ 1999, 2006; Fynbo et al.\ 2003; Le Floc'h et al.\ 2003), and directly measured metallicities (e.g. Stanek et al.\ 2006, Modjaz et al.\ 2008, Kocevski et al.\ 2009). 

Most recently, in Levesque et al.\ (2010a) we published the first results from a spectroscopic survey of LGRB host galaxies, calculating metallicities and other ISM properties for 10 LGRB hosts. A statistical comparison of this sample to the general star-forming galaxy population presents evidence that the ISM properties of these 10 LGRB host galaxies are offset from the general population, and have low-metallicity environments out to $z \sim 1$. This low-metallicity trend agrees with predictions of stellar evolutionary theory (assuming a single-star progenitor model), which postulate that the rapid rotation required to produce a GRB demands low line-driven mass loss rates in the progenitor (e.g. Woosley \& Heger 2006), following the mass loss-metallicity relation of $\dot M_w \propto Z^{0.7}$ (Vink et al.\ 2001). However, our understanding of GRB host environments and progenitors is still evolving, and must strive to accommodate the increasing number of unusual bursts that are detected and observed, as well as potential selection effects introduced by host galaxy luminosities and optical afterglows (e.g. Wolf \& Podsiadlowski 2007, Fynbo et al.\ 2009).

GRB 020819\footnote{While this burst is commonly referred to in the literature as GRB 020819, it is officially designated as GRB 020819B, following the IPN detection of GRB 020819A $\sim$7 hours earlier on 19 Aug 2002.} was originally detected by the High Energy Transient Explorer (HETE), and found to have the energetic properties of a typical long burst, with a duration of $T_{90} \sim 20$s and a peak brightness of $\sim$5 crab (Hurley et al.\ 2002, Vanderspek et al.\ 2002). However, follow-up observations of the burst detected no optical afterglow to a limiting magnitude of $R = 22.2$ and $K' = 19$ at only 9 hours after the burst; this lack of an optical afterglow detection classifies GRB 020819 as a ``dark" burst (Levan et al. 2002, Klose et al.\ 2003). Frail \& Berger (2002) detected a radio afterglow associated with the burst, and Levan et al. (2002) find this position to be coincident with a clearly resolved galaxy at $R \sim 19.8$. Jakobsson et al.\ (2005) later confirmed that this galaxy, at a redshift of $z = 0.41$, was the likely host of GRB 020819, with a chance superposition probability of 0.8\%. The radio afterglow is specifically located on a faint R $\approx$ 24 ``blob" of emission, $\sim$3" from the bright barred spiral host and assumed to be at the same redshift (Jakobsson et al.\ 2005).
 
Klose et al.\ (2003) speculate that the dark nature of GRB 020819 could be due to large amounts of dust extinction in the host. Under this hypothesis, the optical afterglow could have been extincted past the limits of detection by a high host $A_V$. Jakobsson et al.\ (2005) considered this possibility in the context of the host redshift. Using fits to the radio light curve to predict maximum optical fluxes for a variety of afterglow models, they find that a modest amount of extinction - $A_V \sim$ 0.6-1.5 mag - is required to extinguish an optical afterglow with a classical luminosity. However, it is also possible that the optical afterglow of GRB 020819 is undetected due to intrinsically low luminosity, a scenario proposed for other dark bursts (e.g., Fryer et al.\ 1999, De Pasquale et al.\ 2003, Jakobsson et al.\ 2004, Rol et al.\ 2005).

Levesque et al.\ (2010a) include the host galaxy of GRB 051022, another dark burst, in their sample, and find a metallicity of log(O/H) + 12 = 8.62 according to the R$_{23}$ diagnostic of Kewley \& Dopita (2002; see also Kewley \& Ellison 2008). They also find that this host has a modest level of extinction, with E($B-V$) = 0.50 ($A_V = 1.55$). Graham et al.\ (2009) examine the host of GRB 051022 in detail, finding a similarly high metallicity. They speculate that this high metallicity might correlate with GRB 051022's status as a dark burst, and encourage observations of the GRB 020819 host, which with its lower redshift permits a more detailed study of the precise explosion environment for the burst progenitor.
 
In this Letter we investigate the host environment of GRB 020819 in detail. In Section 2, we detail our observations of the host galaxy and explosion site of GRB 020819. We derive a number of ISM properties based on these spectra, and compare these parameters, along with emission line diagnostic ratios, to other LGRB host galaxies, star-forming galaxy samples from the general population, and stellar population synthesis and photoionization models (Section 3). Finally, we consider the implications of this unusual LGRB host environment on our current understanding of LGRB progenitor scenarios (Section 4).

\section{Observations}
We obtained two separate spectra of the GRB 020819 spiral host galaxy, using the Low Resolution Imaging Spectrograph (LRIS; Oke et al.\ 1995) on the Keck I telescope at Mauna Kea. On 2 November 2008 we obtained a spectrum of the host galaxy's nucleus, and on 19 November 2009 we obtained an additional spectrum of the star-forming region associated with the GRB 020819 radio afterglow (Jakobsson et al.\ 2005). We used the long 1" slitmask ($\sim$5.3 kpc at the distance of the host) for both observations along with the 300/5000 grism, the 680 dichroic, and the 400/8500 grating at a central wavelength of 8100\AA. In 2008 the slit was centered on a nearby bright star and turned to the position angle that would place both the star and the nucleus of the GRB 020819 host on the slit (PA = 345.88$^{\circ}$). In 2009, the slit was positioned such that both the host galaxy nucleus and the ``blob" designated as the explosion site by Jakobsson et al.\ (2005) were positioned on the slit (PA = 9.5$^{\circ}$). As a result, the observations were not taken at the parallactic angle.

The data were reduced using IRAF\footnotemark.\footnotetext{IRAF is distributed by NOAO, which is operated by AURA, Inc., under cooperative agreement with the NSF.} We used the IRAF tasks \texttt{lrisbias} and \texttt{multi2simple} distributed by the W. M. Keck Observatories to subtract overscan from the images. The spectra were extracted using the \texttt{apall} task in the \texttt{kpnoslit} package. Wavelength and flatfield calibrations were performed based on our observations of internal lamp flat fields and Hg, Ne, Ar, Cd, and Zn comparison lamp spectra. Flux calibration was performed using our observations of the spectrophotometric standard GD 248 (Oke 1990) on both nights.

From our 2008 observations, we detected [OII]$\lambda$3727, H$\beta$, [OIII]$\lambda$5007, H$\alpha$, and [NII]$\lambda\lambda$6548,6584 features in emission for the host galaxy nucleus. In 2009, problems with the response of the blue side CCD unfortunately prevented detection of H$\beta$ and [OIII]$\lambda$5007 emission features in the explosion site spectra; however, [OII]$\lambda$3727, H$\alpha$, and [NII]$\lambda\lambda$6548,6584 emission features were detected at a redshift of $z = 0.41$, confirming its association with the bright spiral host. Emission line fluxes were determined using the IRAF task \texttt{splot} in the \texttt{kpnoslit} package to fit Gaussians to the line profiles. The detected emission lines for the nucleus and explosion site spectra are shown in Figure 1.

\section{Analyses}
\subsection{ISM Properties of the GRB 020819 host galaxy}
We determined E($B-V$) in the direction of GRB 020819 based on H$\alpha$ and H$\beta$ fluxes measured in the nucleus spectrum. We adopt the Cardelli et al.\ (1989) reddening law with $R_V = 3.1$, a Balmer decrement of H$\alpha$/H$\beta$ = 2.87 (following Osterbrock 1989 for case B recombination), and the wavelength-dependent constant k(H$\alpha$) = 2.535 (Cardelli et al.\ 1989). We find a total line-of-sight E($B-V$) = 0.71 mag for the nucleus of the GRB 020819 host galaxy; when Galactic extinction E($B-V$) = 0.07 in the direction of the host (Schlegel et al.\ 1998) is accounted for, this suggests a host E($B-V$) = 0.64, or a host $A_V = 1.98$. In the absence of an H$\beta$ detection in the ``blob" spectrum, we cannot determine an E($B-V$) for the explosion site. However, our $A_V$ exceeds the required host extinction of $A_V \approx 0.6 - 1.5$ mag proposed by Jakobsson et al.\ (2005) to account for the absence of an optical afterglow. We adopt E($B-V$) = 0.71 as an approximation for the amount of extinction present at the explosion site.

The metallicity diagnostics presented in Kewley \& Dopita (2002) and Kewley \& Ellison (2008) use strong optical emission line ratios to determine metallicities and ionization parameters based on equations derived from photoionization models. Using the Kewley \& Dopita (2002) polynomial relation between [NII]$\lambda$6584/[OII]$\lambda$3727 and metallicity, we find log(O/H) + 12 = 9.0 $\pm$ 0.1 for the nucleus of the GBR 020819 host galaxy. We also adopt the [NII]$\lambda$6584/H$\alpha$ diagnostic relation presented in Pettini \& Pagel (2004), a diagnostic independent of extinction effects, and find a metallicity of log(O/H) + 12 =  8.8 $\pm$ 0.1. As this calibration is based on an empirical fit to electron-temperature based metallicities, it is known to be systematically offset from calibrations derived from theoretical photoionization models by $\sim$0.2-0.3 dex; this is in agreement with the metallicity discrepancy seen here (e.g. Kewley \& Ellison 2008 and references therein).

From our spectrum of the explosion site, we similarly find log(O/H) + 12 = 9.0 $\pm$ 0.1 adopting the Kewley \& Dopita (2002) [NII]/[OII] relation and log(O/H) + 12 = 8.7 $\pm$ 0.1 adopting the Pettini \& Pagel (2004) [[NII]$\lambda$6584/H$\alpha$ diagnostic. For these abundance determinations we adopt the E($B-V$) = 0.71 associated with the nucleus of the host. At lower E($B-V$) the metallicity derived from the Kewley \& Dopita (2002) [NII]/[OII] relation increases (and vice versa), but the Pettini \& Pagel (2004) [NII]/H$\alpha$ metallicity stays constant; as a result we use this metallicity diagnostic for comparison throughout the remainder of this Letter. The metallicity measured at the explosion site is identical to that measured at the nucleus, to within the errors.

Adopting $R$ magnitudes from Jakobsson et al.\ (2005), and $M^*_R = -21.57$ from Brown et al.\ (2001), we find a luminosity of $\sim2L*$ for the host galaxy and $\sim0.05L*$ for the explosion site. We determine a young stellar population age of 7.8 $\pm$ 0.9 Myr for the nucleus of the GRB 020819 host galaxy, based on a relation with the equivalent width of the H$\beta$ emission line from Levesque et al.\ (2010a, eqn. 2); however, we are unable to apply this age determination to the explosion site due to our lack of an H$\beta$ emission line detection. For the nucleus we find an extinction-corrected star formation rate of 23.6 M$_{\odot}$/yr based on the flux of the H$\alpha$ line (Kennicutt 1998); for the explosion site we find 10.2 M$_{\odot}$/yr.

The ISM properties of the GRB 020819 host galaxy nucleus and explosion site are summarized in Table 1.

\subsection{Comparison With LGRB Host Galaxies}
In Figure 2 we compare GRB 020819 to six other $z < 1$ LGRB host galaxies from Levesque et al.\ (2010a) on a plot of young stellar population age vs. metallicity. For comparison we also include a sample of 7 $z < 0.1$ metal-poor galaxies from Brown et al.\ (2008) and a sample of blue compact galaxies from Kong \& Cheng (2002; for further discussion of these comparison samples see Levesque et al.\ 2010a). It is evident that both the metallicity and age of the GRB 020819 host galaxy set it apart from the current sample of LGRB host galaxies. Adopting the Pettini \& Pagel (2004) [NII]/H$\alpha$ metallicity diagnostic, the explosion site metallicity of log(O/H) + 12 = 8.7 $\pm$ 0.1 is considerably higher than the mean LGRB host metallicity of log(O/H) + 12 = 8.1 $\pm$ 0.2 derived using the same diagnostic. Similarly, the young stellar population age of 7.8 $\pm$ 0.9 Myr determined for the GRB 020819 host galaxy nucleus is older than the mean age of 5.2 $\pm$ 0.15 Myr calculated for the young stellar populations of LGRB host galaxies. Both of these disparities suggest that the progenitor of GRB 020819 formed in a markedly different environment than that seen in other LGRB host galaxies.

In Figure 3 we place our GRB 020819 host observations on the [NII]/H$\alpha$ vs. [OIII]/H$\beta$ emission line ratio diagnostic diagram of Baldwin et al.\ (1981; top) and the [NII]/[OII] vs. [OIII]/[OII] diagnostic diagram of Dopita et al.\ (2000; bottom), comparing its placement to the six LGRB hosts from Levesque et al.\ (2010a). Also included on these diagrams for comparison is a sample of 60920 $z < 0.1$ star-forming galaxies from Kewley et al.\ (2006), as well as the sample of metal-poor galaxies from Brown et al.\ (2008). Finally, we include the new grid of stellar population synthesis and photoionization models published in Levesque et al.\ (2010b), to provide an independent comparison of metallicity and ionization parameter. We can see that emission line diagnostic ratios determined for both the GRB 020819 host galaxy nucleus and explosion site are distinct from the LGRB host galaxies of Levesque et al.\ (2010a). In Figure 3 (bottom) we do see one LGRB host galaxy that appears to have a similarly high metallicity; however, this data point represents the host galaxy of GRB 031203, which has shown evidence of AGN activity that may contaminate emission-line-based determinations of its ISM properties (Levesque et al.\ 2010a). The GRB 020819 host galaxy shows no similar signs of such activity; in both diagnostic diagrams the GRB 020819 host appears to be very similar to a typical star-forming SDSS galaxy, in marked contrast to the other LGRB host galaxies.

\section{Discussion}
Our spectroscopic observations and metallicity diagnostics have demonstrated that GRB 020819 did {\it not} occur in a low-metallicity region of the spiral host, but rather that the progenitor formed and evolved in a host environment with a super-solar metallicity. This is further supported by examining the position of the GRB 020819 nucleus and explosion site on the emission line diagnostic diagrams of Figure 3. This high metallicity sets GRB 020819 apart from all other LGRB host galaxies examined to date - while some hosts appear to have solar or near-solar metallicities based on afterglow spectra, the relationship between afterglow absorption metallicities and emission-line metallicities has not been examined and these values may not be directly comparable. It is also possible that higher-metallicity host galaxies might be missing from the current sample due to selection effects; the size and nature of this bias is addressed in detail in Graham et al.\ in prep.

Due to the young lifetimes ($\le$ 10 Myr, Woosley et al.\ 2002) associated with the assumed massive star progenitors of LGRBs, we can take the metallicities of their host environments to be representative of the progenitor star metallicities. The high-metallicity ISM environment that produced GRB 020819 is therefore particularly surprising in the context of our current understanding of progenitor scenarios and evolution for LGRBs. It has been suggested that low metallicities are required to produce the rapidly-rotating progenitors and relativistic explosions associated with LGRBs (e.g. Vink et al.\ 2001, Meynet \& Maeder 2005, Woosley \& Heger 2006) and the general low metallicity trend seen in LGRB host galaxies supports this claim (e.g. Modjaz et al.\ 2008, Kocevski et al.\ 2009, Levesque et al.\ 2010a). However, the exact nature of this correlation between metallicity and GRB progenitors remains unclear. The recent Type Ic supernova SN 2009bb was found to have a central-engine-driven relativistic component similar to that associated with LGRBs, although no accompany gamma-ray trigger was associated with the event (Soderberg et al.\ 2009). The host environment of this supernova was found to have a very high metallicity (Levesque et al.\ 2010c), contradicting the supposition that engine-driven relativistic supernovae could only be produced in low-metallicity environments.

While its metallicity is not as high by comparison, the only other LGRB host galaxy with a $\sim Z_{\odot}$ metallicity is GRB 051022. Graham et al.\ (2009) point out that GRB 051022 is, like GRB 020819, a ``dark" burst, and postulate that the higher metallicity seen in GRB 051022 may be shared by the GRB 020819 host environment, a speculation that has been confirmed by these results. The question of what role high metallicity might play in the production of ``dark" bursts is an intriguing one, and requires a detailed comparison of the host environments and energetic properties of these two bursts (see Graham et al., in prep, for a comparison of GRB 020819 and GRB 051022 as well as new HST observations of the GRB 051022 host galaxy). While both host galaxies include a moderate amount of extinction ($A_V$ = 1.55 for GRB 051022, $A_V$ = 1.98 for GRB 020819), we cannot rule out the possibility that direct effects of high metallicity environments on progenitor evolution might be responsible for the ``dark" nature of these bursts. For example, the enhanced mass loss rates associated with higher metallicities could potentially contribute to large amount of circumburst extinction, an artifact of larger amounts of mass lost during the progenitor's lifetime (e.g., Vink et al.\ 2001). This is particularly intriguing if anisotropies in stellar winds and mass loss are considered for a single-star progenitor model. Polar mass ejections remove much less angular momentum than equatorial ejections (Maeder 2002). A polar mass loss mechanism could both permit the production of GRB progenitors at higher metallicities by allowing higher rotation rates to be maintained, {\it and} increase the circumburst extinction at the poles where the GRB itself is produced, leading to increased extinction of the optical afterglow. However, the full implications that GRB 020819 and its unusual host environment have for our understanding of ``dark" bursts and their progenitor evolution still remain to be explored. 

We gratefully acknowledge the hospitality of W. M. Keck Observatories in Hawaii, particularly the assistance of Greg Wirth. This paper made use of data from the Gamma-Ray Burst Coordinates Network (GCN) circulars. The authors wish to recognize and acknowledge the very significant cultural role and reverence that the summit of Mauna Kea has always had within the indigenous Hawaiian community. We are fortunate to have the opportunity to conduct observations from this sacred mountain. E. Levesque's participation was partially made possible by a Ford Foundation Predoctoral Fellowship. L. Kewley and E. Levesque gratefully acknowledge support by NSF EARLY CAREER AWARD AST07-48559.

\begin{deluxetable}{l c c c c c c c c c c c c}
\tabletypesize{\scriptsize}
\tablewidth{0pc}
\tablenum{1}
\tablecolumns{13}
\tablecaption{\label{tab:gals} GRB 020819 Host Environment ISM Properties}
\tablehead{
\colhead{}
&\colhead{}
&\multicolumn{5}{c}{Measured Emission Line Fluxes\tablenotemark{a}}
&\colhead{}
&\multicolumn{5}{c}{Derived ISM Properties} \\ \cline{3-7} \cline {9-13}
\colhead{Region}
&\colhead{$z$}
&\colhead{[OII]}
&\colhead{H$\beta$}
&\colhead{[OIII]}
&\colhead{H$\alpha$}
&\colhead{[NII]}
&\colhead{}
&\colhead{E($B-V$)\tablenotemark{b}}
&\multicolumn{2}{c}{log(O/H) + 12}
&\colhead{Age}
&\colhead{SFR} \\ \cline{10-11}
\colhead{}
&\colhead{}
&\colhead{3727\AA}
&\colhead{}
&\colhead{(5007\AA}
&\colhead{}
&\colhead{6584\AA}
&\colhead{}
&\colhead{(mag)}
&\colhead{[NII]/[OII]}
&\colhead{[NII]/H$\alpha$}
&\colhead{(Myr)}
&\colhead{(M$_{\odot}$ yr$^{-1}$)}
}
\startdata
Nucleus &0.411 &2.41 &1.66 &0.86 &9.62 &4.10 & &0.71 &9.0$\pm$ 0.1 &8.8$\pm$ 0.1 &7.8 $\pm$ 0.9 &23.6 \\
Explosion site & 0.411 &0.92 &\nodata &\nodata &4.14 &1.51 & &\nodata\tablenotemark{c} &9.0$\pm$ 0.1 &8.7$\pm$ 0.1 &\nodata\tablenotemark{c} &10.2 \\
\enddata	      
\tablenotetext{a}{Uncorrected fluxes in units of 10$^{-16}$ ergs cm$^2$ s$^{-1}$ \AA$^{-1}$. The fluxes have a systematic error of $\sim$10\% ($\pm$5\%) incurred by errors in the flux calibration; this dominates any statistical errors.}
\tablenotetext{b}{Total color excess in the direction of the galaxy, used to correct for effects of both Galactic and host extinction.}
\tablenotetext{c}{Could not be calculated due to lack of an H$\beta$ emission detection in the spectrum as a result of detector problems.}
\end{deluxetable}

\begin{figure}
\epsscale{0.45}
\plotone{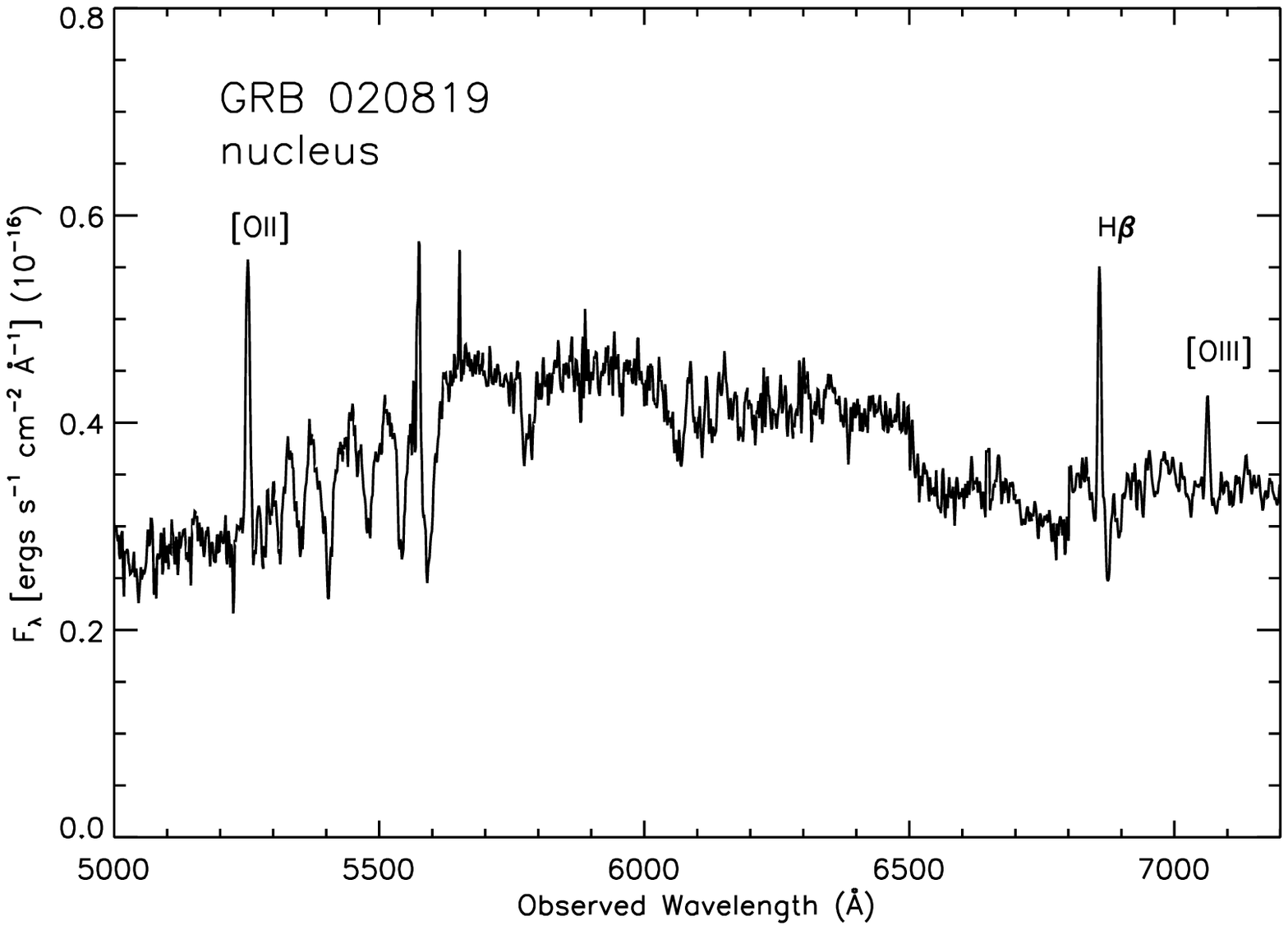}
\plotone{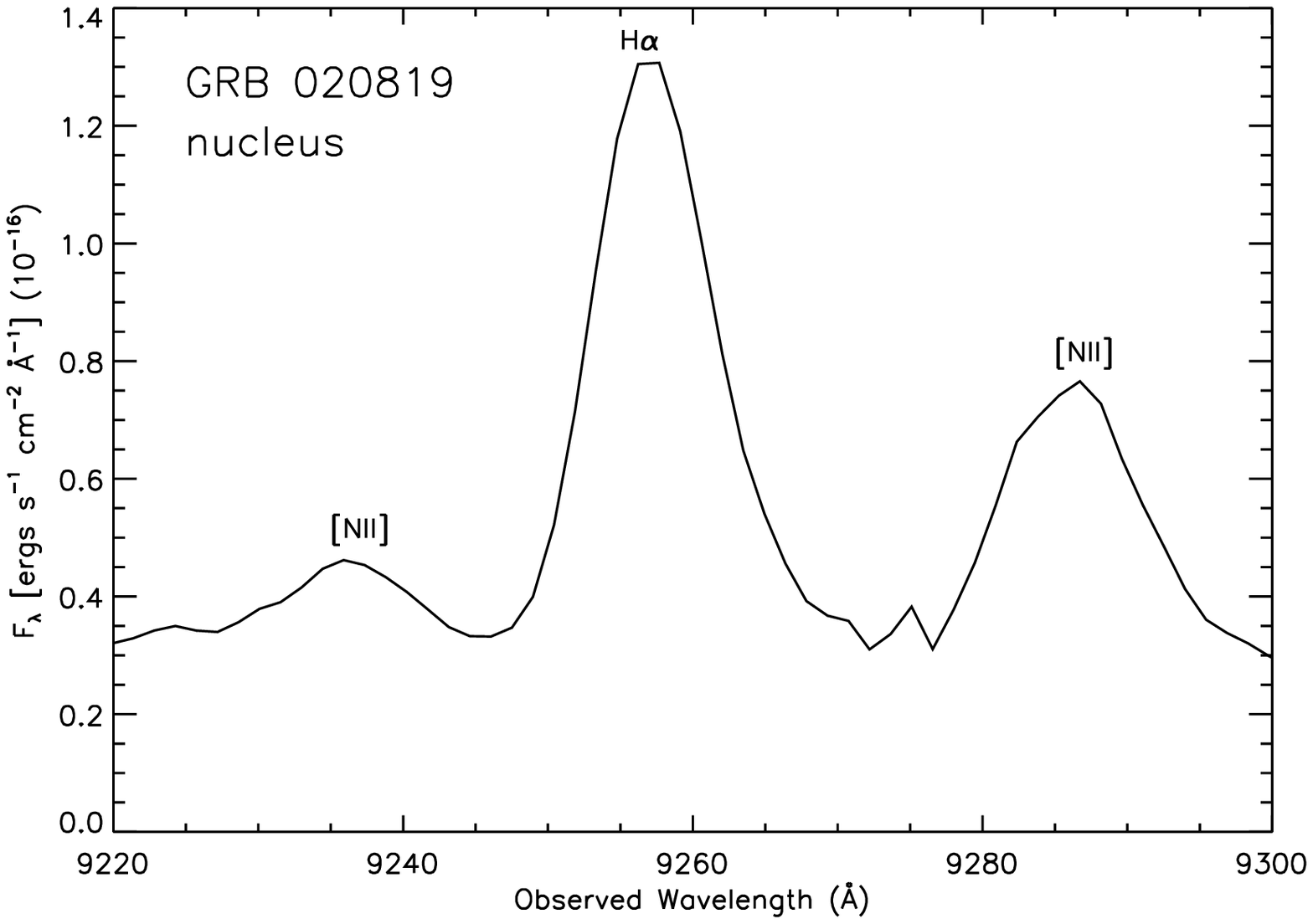}
\plotone{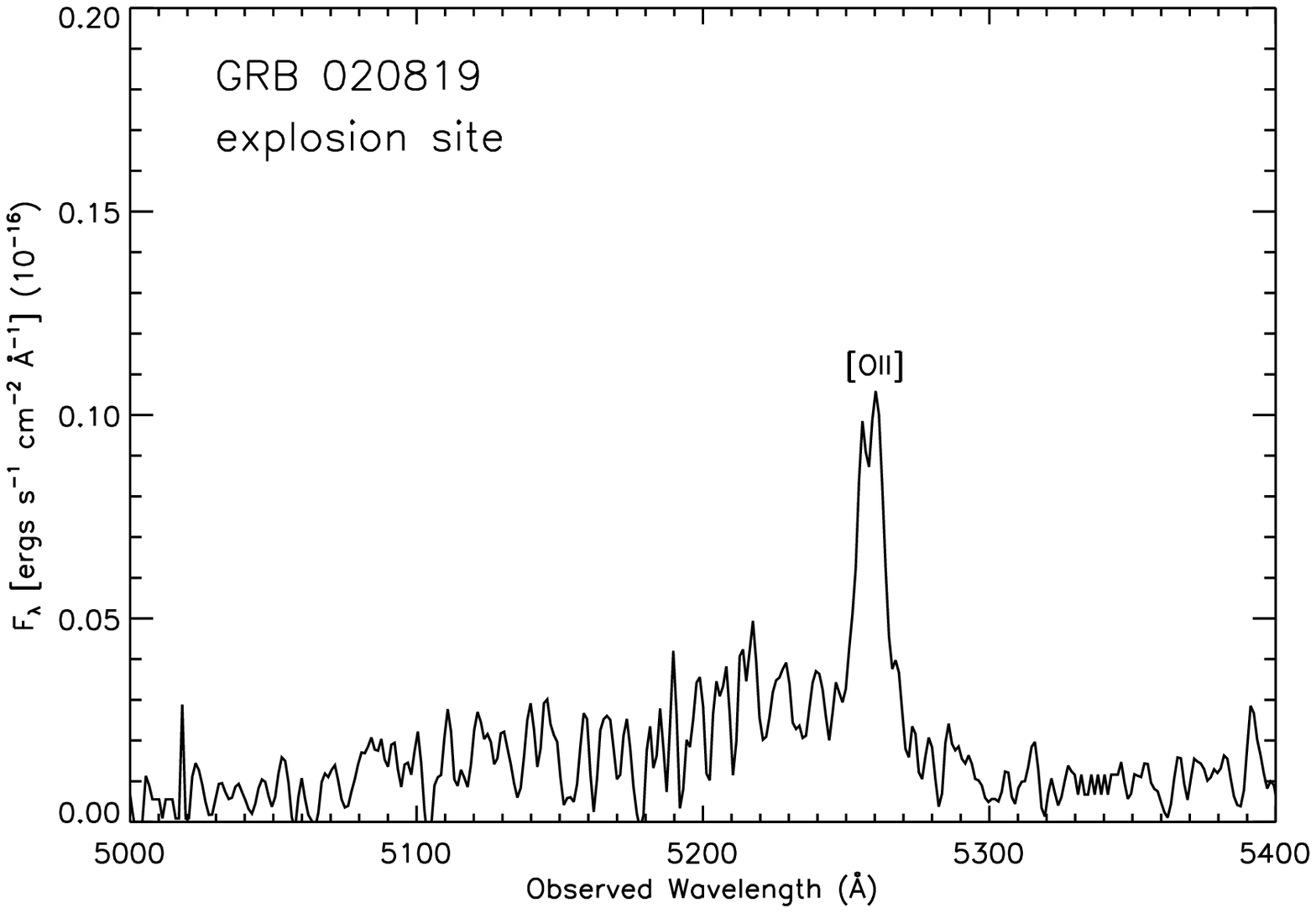}
\plotone{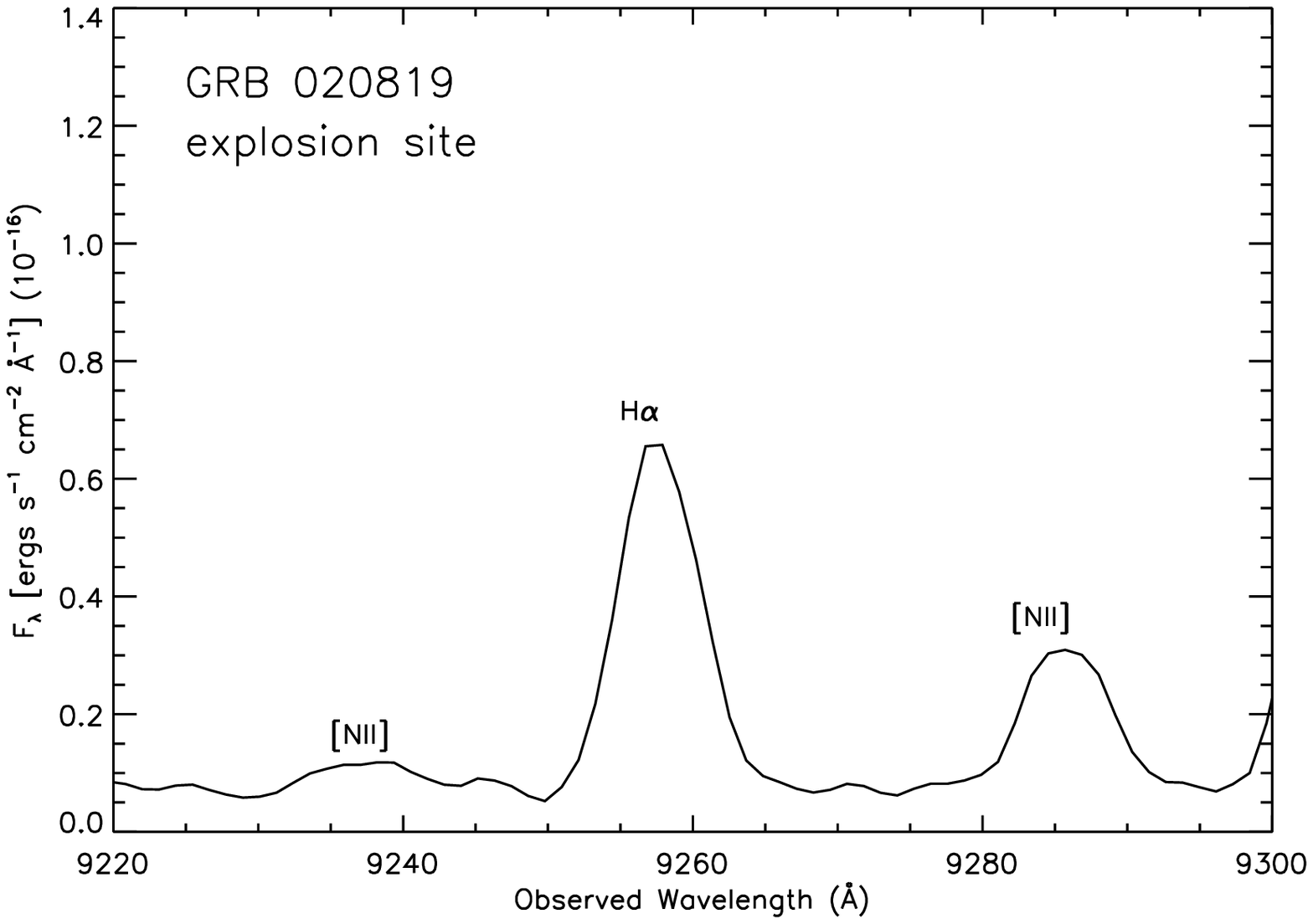}
\caption{Spectra of the GRB 020819 host galaxy nucleus (top; November 2008) and explosion site (bottom; November 2009), showing the rest-frame optical emission lines detected in the blue (left) and red (right).}
\end{figure}

\begin{figure}
\epsscale{0.8}
\plotone{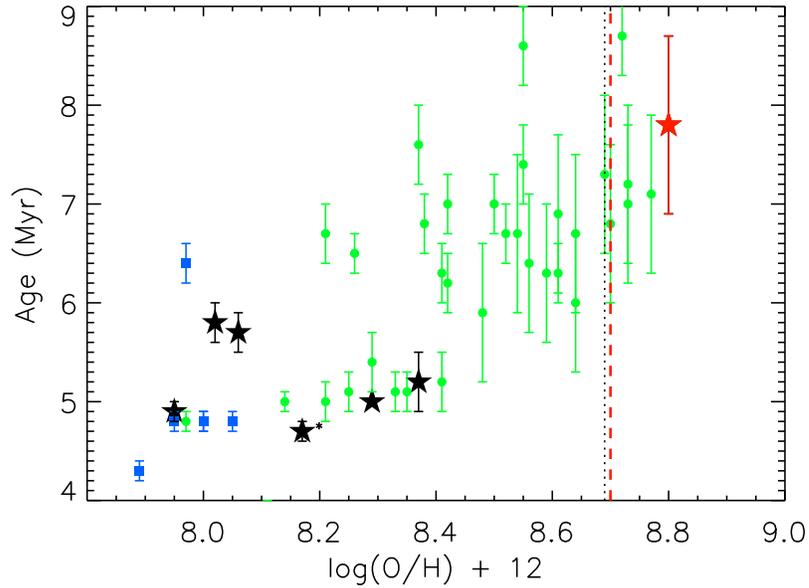}
\caption{Comparison of age vs. metallicity for LGRB host galaxy metallicities. Included are six LGRB host metallicities from Levesque et al.\ (2010a), adopting the Pettini \& Pagel (2004) [NII]/H$\alpha$ emission line diagnostic calibration (black stars) (here and in Figure 3, the host galaxy marked with a ``*" is the host of GRB 031203, which is thought to be contaminated by AGN activity).The central bulge of the GRB 020819 host is shown as a red star, and the metallicity of the GRB 020819 explosion site is plotted as a red dashed line. Also included for comparison are a sample of metal-poor galaxies from Brown et al.\ (2008; blue squares) and a sample of blue compact galaxies from Kong \& Cheng (2002; green circles). The Asplund et al.\ (2005) solar metallicity log(O/H) + 12 = 8.69 is plotted as a black dotted line. We can see that the host environment of GRB 020819 has a much higher metallicity than the other LGRB hosts included in this sample, and shows much more similarity to the blue compact galaxy sample as opposed to the metal-poor galaxies.} 
\end{figure}

\begin{figure}
\epsscale{0.7}
\plotone{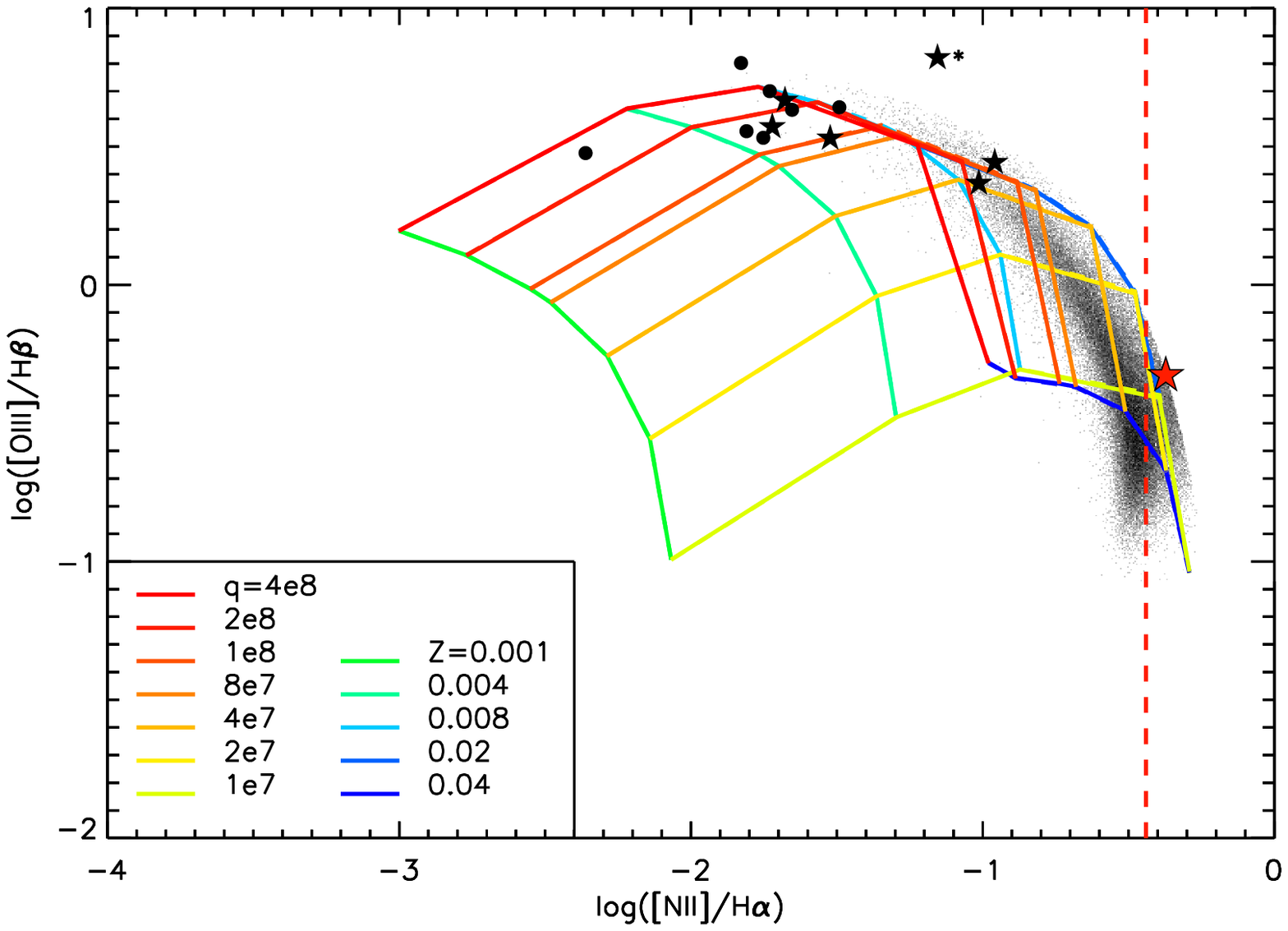}
\plotone{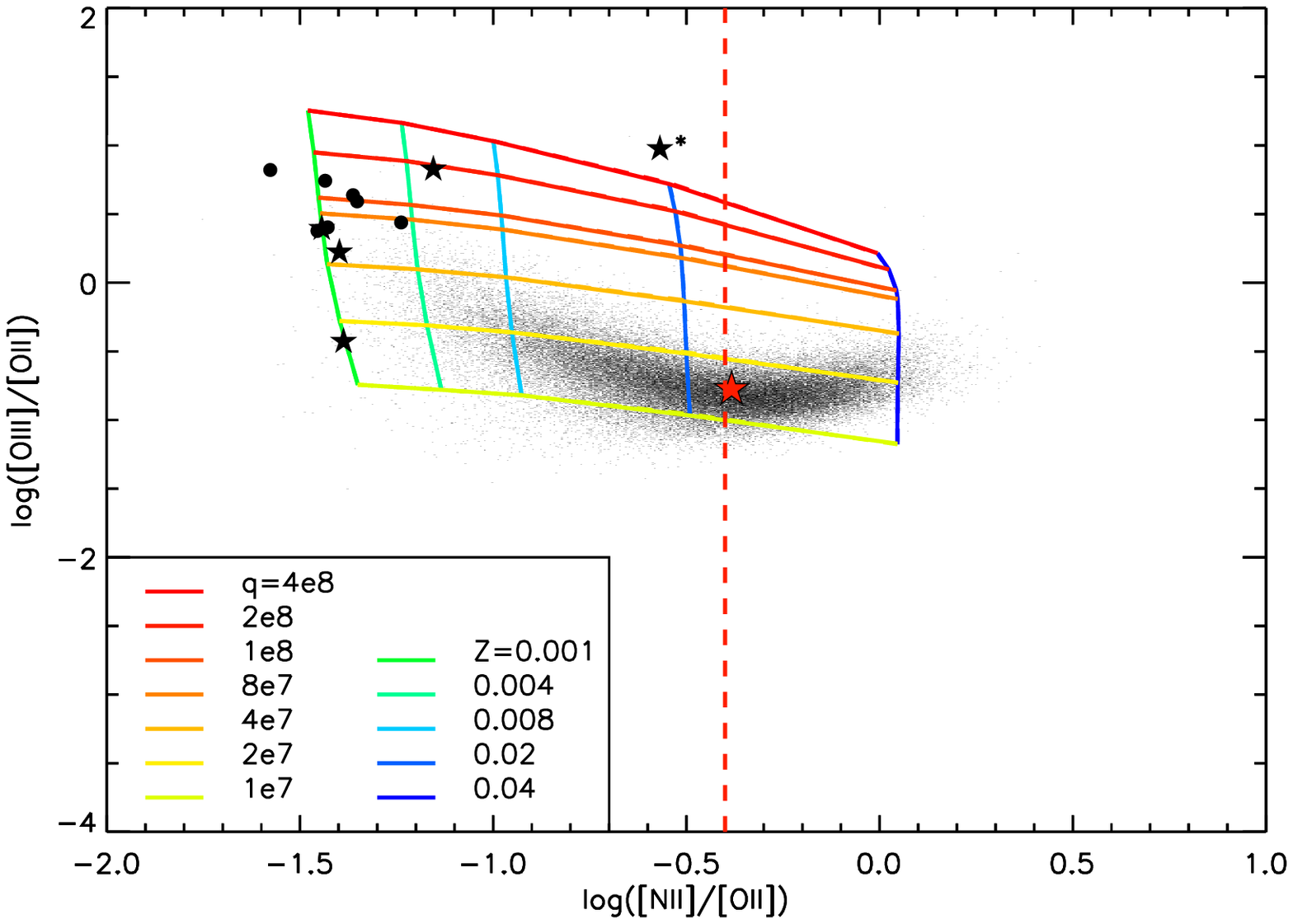}
\caption{Placement of the GRB 020819 host galaxy nucleus (red star) and explosion site (red dashed line) on the [NII]/H$\alpha$ vs. [OIII]/H$\beta$ (top) and [NII]/[OII] vs. [OIII]/[OII] (bottom) diagnostic diagrams. Included for comparison are six $z < 1$ LGRB host galaxies from Levesque et al.\ (2010a; black stars), a sample of 60920 $z < 0.1$ star-forming SDSS galaxies (Kewley et al.\ 2006; points), seven $z < 0.1$ metal-poor galaxies from Brown et al.\ (2008; black dots), and the stellar population synthesis and photoionization models of Levesque et al.\ (2010b; colored grids), plotted with lines of constant metallicity and ionization parameter. The emission line ratios measured for the GRB 020819 host galaxy nucleus and explosion site are distinct from the other LGRB hosts included in the sample, and comparable to a typical star-forming SDSS galaxy.}
\end{figure}


\begin{references}
\reference {} Asplund, M., Grevesse, N., \& Sauval, A. J. 2005, in Astronomical Society of the Pacific Conference Series, Vol. 336, Cosmic Abundances as Records of Stellar Evolution and Nucleosynthesis, ed. T. G. Barnes, III \& F. N. Bash, 25
\reference {} Brown, W., Geller, M. J., Fabricant, D. G., \& Kurtz, M. J. 2001, AJ, 122, 714
\reference {} Brown, W., Kewley, L. J., \& Geller, M. J. 2008, AJ, 135, 92
\reference {} Cardelli, J. A., Clayton, G. C., \& Mathis, J. S. 1989, ApJ, 345, 245
\reference {} Christensen, L., Vreeswijk, P. M., Sollerman, J., Th\"{o}ne, C. C., Le Floc'h, E., \& Wiersema, K. 2008, A\&A, 490, 45
\reference {} De Pasquale, M. et al.\ 2003, ApJ, 592, 1018
\reference {} Frail, D. A., \& Berger, E. 2003, GCN Circ. 1842
\reference {} Fruchter, A. S. et al.\ 1999, ApJ, 519, L13
\reference {} Fruchter, A. S. et al.\ 2006, Nature, 441, 463
\reference {} Fryer, C. L.,Woosley, S. E., \& Hartmann,D.H. 1999, ApJ, 526, 152
\reference {} Fynbo, J. P. U. et al.\ 2003, A\&A, 406, 63
\reference {} Fynbo, J. P. U. et al.\ 2009, ApJS, 185, 526
\reference {} Galama, T. J. et al. 1998, Nature, 395, 670
\reference {} Graham, J. F., Fruchter, A. S., Kewley, L. J., Levesque, E. M., Levan, A. J., Tanvir, N. R., Reichart, D. E., \& Nysewander, M. 2009, in Gamma-Ray Burst: Sixth Huntsville Symposium, American Institute of Physics Conference Proceedings 1133, p. 269
\reference {} Graham, J. F. et al. 2010, in prep
\reference {} Hurley, K., et al. 2002, GCN Circ. 1507
\reference {} Jakobsson, P., Hjorth, J., Fynbo, J. P. U., Watson, D., Pedersen, K., Bjornsson, G., \& Gorosabel, J. 2004, ApJ, 617, L21
\reference {} Jakobsson, P. et al.\ 2005, ApJ, 629, 45
\reference {} Kennicutt, R. C. 1998, ARA\&A, 36, 189
\reference {} Kewley, L. J. \& Dopita, M. A. 2002, ApJS, 142, 35
\reference {} Kewley, L. J., Groves, B., Kauffman, G., \& Heckman, T. 2006, MNRAS, 372, 961
\reference {} Kewley, L. J., Ellison, S. L. 2008, ApJ, 681, 1183
\reference {} Klose, S. et al.\ 2003, ApJ, 592, 1025
\reference {} Kobulnicky, H. A. \& Kewley, L. J. 2004, ApJ, 617, 24
\reference {} Kocevski, D., West, A. A., \& Modjaz, M. 2009, ApJ, 702, 377
\reference {} Kong, X. \& Cheng, F. Z. 2002, A\&A, 389, 845
\reference {} Le Floc'h, E. et al.\ 2003, A\&A, 400, 499
\reference {} Levan, A., et al. 2002, GCN Circ. 1844
\reference {} Levesque, E. M., Berger, E., Kewley, L. J., \& Bagley, M. M. 2010a, AJ, in press
\reference {} Levesque, E. M., Kewley, L. J., \& Larson, K. 2010b, AJ, in press
\reference {} Levesque, E. M. et al.\ 2010c, ApJL, in press
\reference {} Maeder, A. 2002, A\&A, 392, 575
\reference {} Meynet, G. \& Maeder, A. 2005, A\&A, 429, 581
\reference {} Modjaz, M., Kewley, L. J., Kirshner, R. P., Stanek, K. Z., Challis, P., Garnavich, P. M., Greene, J. E., Kelly, P. L., Prieto, J. L. 2008, AJ, 135, 1136
\reference {} Oke, J. B. 1990, AJ, 99, 1621
\reference {} Oke, J. B. et al.\ 1995, PASP, 107, 375
\reference {} Osterbrock, D.1989, Astrophysics of gaseous nebulae and active galactic nuclei (University Science Books)
\reference {} Pettini, M. \& Pagel, B. E. J. 2004, MNRAS, 348, 59
\reference {} Rol, E., Wijers, R. A. M. J., Kouveliotou, C., Kaper, L., \& Kaneko, Y. 2005, ApJ, 624, 868
\reference {} Schlegel, D. J., Finkbeiner, D. P., \& Davis, M. 1998, ApJ, 500, 525
\reference {} Soderberg, A. M. et al.\ 2010, Nature, in press
\reference {} Stanek, K. Z. et al.\ 2003, ApJ, 591, L17
\reference {} Stanek, K. Z., et al.\ 2006, Acta Astron. 56, 333
\reference {} Vanderspek, R., et al. 2002, GCN Circ. 1508
\reference {} Vink, J. S., de Koter, A., \& Lamers, H. J. G. L. M. 2001, A\&A, 369, 574
\reference {} Wolf, C. \& Podsiadlowski, P. 2007, MNRAS, 375, 1049
\reference {} Woosley, S. E. 1993, ApJ, 405, 273
\reference {} Woosley, S. E., Heger, A., \& Weaver, T. A. 2002, RMP, 74, 1015
\reference {} Woosley, S. E. \& Bloom, J. S. 2006, ARA\&A, 44, 507
\reference {} Woosley, S. E. \& Heger, A. 2006, ApJ, 637, 914
\end{references}
\end{document}